\documentclass[twoside,a4paper,11pt]{sea10}
\usepackage{graphicx}
\usepackage{hyperref}
\usepackage{amssymb,amsmath}
\begin{document}
\pagenumbering{arabic}
\pagestyle{myheadings}
\thispagestyle{empty}
{\flushleft\includegraphics[width=\textwidth,bb=58 650 590 680]{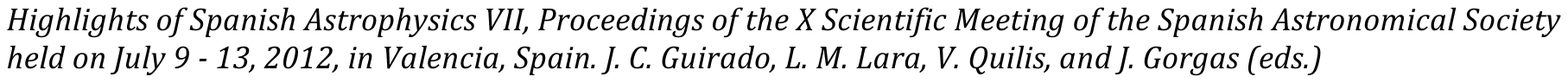}}
\vspace*{0.2cm}
\begin{flushleft}
{\bf {\LARGE
%
Measuring vector magnetic fields in solar prominences  %
}\\
\vspace*{1cm}
%
D.\ Orozco Su\'arez$^{1,2}$,
A.\ Asensio Ramos$^{1,2}$, 
and 
J.\ Trujillo Bueno$^{1,2,3}$
%
}\\
\vspace*{0.5cm}
%
$^{1}$
Instituto de Astrof\'isica de Canarias, E-38205 La Laguna, Tenerife, Spain\\
$^{2}$
Dpto.\ de Astrof\'isica, Universidad de La Laguna, E-38206 La Laguna, Tenerife, Spain\\
$^{3}$
Consejo Superior de Investigaciones Cient\'ificas, Spain

%
\end{flushleft}
%
\markboth{
Measuring vector magnetic fields in solar prominences
}{ 
%
Orozco Su\'arez, Asensio Ramos, and Trujillo Bueno
%
}
\thispagestyle{empty}
\vspace*{0.4cm}
\begin{minipage}[l]{0.09\textwidth}
\ 
\end{minipage}
\begin{minipage}[r]{0.9\textwidth}
\vspace{1cm}
\section*{Abstract}{\small
%
We present spectropolarimetric observations in the He~{\small I}~1083.0 nm multiplet of a quiescent, hedgerow solar prominence. The data were taken with the Tenerife Infrared Polarimeter attached to the German Vacuum Tower Telescope at the Observatorio del Teide (Tenerife; Canary Islands; Spain). The observed He~{\small I} circular and linear polarization signals are dominated by the Zeeman effect and by atomic level polarization and the Hanle effect, respectively. These observables are sensitive to the strength and orientation of the magnetic field vector at each spatial point of the field of view. We determine the magnetic field vector of the prominence by applying the HAZEL inversion code to the observed Stokes profiles. We briefly discuss the retrieved magnetic field vector configuration.
%
\normalsize}
\end{minipage}

%
%
%
\section{Introduction \label{intro}}

Prominences, or filaments when seen against the solar disk, are cool, dense, magnetized structures of 10$^4$~K plasma embedded in the 10$^6$~K solar corona and located above the so-called polarity inversion line or filament channel (see \cite{labrosse,mackay} for reviews). Magnetic field measurements in prominences have been made since the 1960's using the Zeeman effect and since the 1980's using the Hanle effect \cite{lopezariste}. As a result, we have a fairly clear picture of the global magnetic properties of solar prominences: the magnetic field is rather uniform along the prominence and has strengths of the order of 10~G. The orientation of the field vector with respect the solar surface is nearly horizontal, forming an acute angle of about 35$^{\mathrm{o}}$ with the prominence long axis (e.g., \cite{1994SoPh..154..231B}). The configuration seems to be different for polar crown prominences where the field is found to be nearly perpendicular to the solar surface \cite{merenda}. However, the current understanding about the three dimensional configuration of the magnetic field and about its small-scale structuring is still insufficient in order to verify theoretical models. There are very few research papers providing a glimpse on the possible spatial distribution of the magnetic field vector in quiescent prominences (e.g., \cite{casini,merenda}). Of particular interest is an investigation using the Hanle effect on the He~{\small I}~D3 line \cite{casini}. Their results were unexpected since they found that the field was organized in patches where the field increases locally up to 80~G. Thus, there is a clear need to obtain vector-field maps of solar prominences on a regular basis to advance our knowledge in their magnetic field vector configuration and in characterizing local magnetic field variations, if any, and also to validate or reject proposed theoretical models.

The He~{\small I}~1083.0~nm multiplet is of particular interest to measure the magnetic field vector in prominences and filaments \cite{2009AA...501.1113K,Kuckein1,Kuckein2,merenda,2007ASPC..368..347M,2002Natur.415..403T}, and spicules \cite{2010ApJ...708.1579C,2005ApJ...619L.191T}. The reason is that it can be clearly seen in emission in off-limb prominences (e.g., \cite{2002Natur.415..403T}) and that it is sensitive to the joint action of atomic level polarization (i.e., population imbalances and quantum coherence among the magnetic sublevels, generated by anisotropic radiation pumping) and the Hanle (modification of the
atomic level polarization due to the presence of a magnetic field) and Zeeman effects \cite{2007ApJ...655..642T,2002Natur.415..403T}. The later makes the He~{\small I}~1083.0~nm multiplet sensitive to a wide range of field strengths from dG (Hanle) to kG (Zeeman). An user-friendly diagnostic tool called HAZEL\footnote{From HAnle and ZEeman Light. Visit http://www.iac.es/proyecto/magnetism/pages/codes/hazel.php for getting more information about the code.} is
available for modeling and interpreting the He~{\small I}~1083.0~nm triplet polarization signals, in order to determine the strength, inclination and azimuth of the magnetic field vector \cite{2008ApJ...683..542A}. This code has already been applied to observational data \cite{2010MmSAI..81..625A,2010ApJ...708.1579C,marian}.

Here we present a preliminary analysis of ground-based spectropolarimetric observations of the He~{\small I} triplet taken in a quiescent, hedgerow prominence.  The observations, analyzed with the HAZEL code, have allowed us to construct a two dimensional map of the magnetic field vector of the prominence with a spatial resolution of about 1$^{\prime\prime}$. 

\section{Observations, inference method, and preliminary results}

\begin{figure}[t]
\center
\includegraphics[scale=0.6]{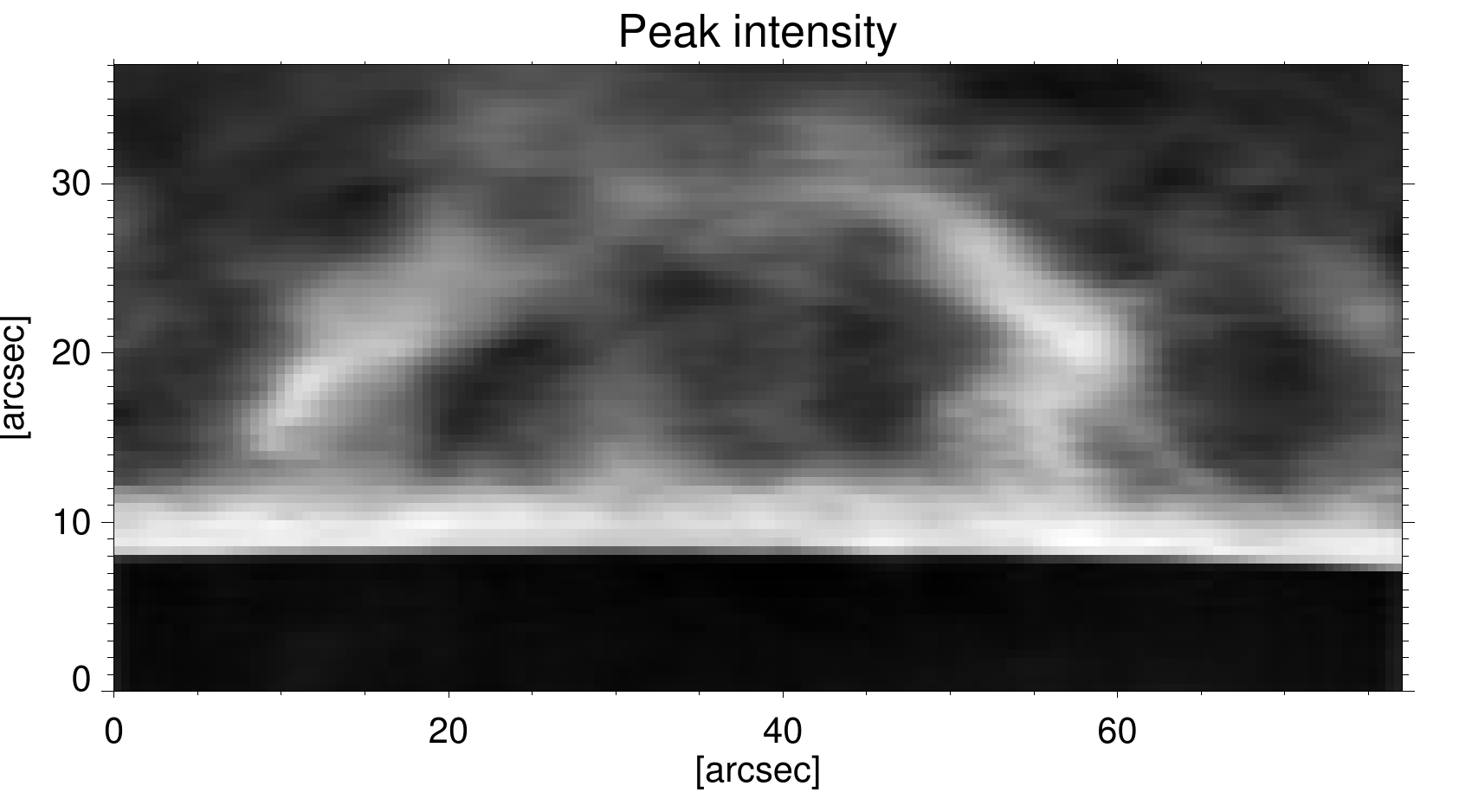}
\caption{\label{fig1} Peak intensity of the He~{\small I}~1083.0~nm emission profile. The prominence is seen as a bright structure against a dark background. The bottom, dark part corresponds to the solar limb.}
\end{figure}

\begin{figure}[t]
\center
\includegraphics[scale=0.6]{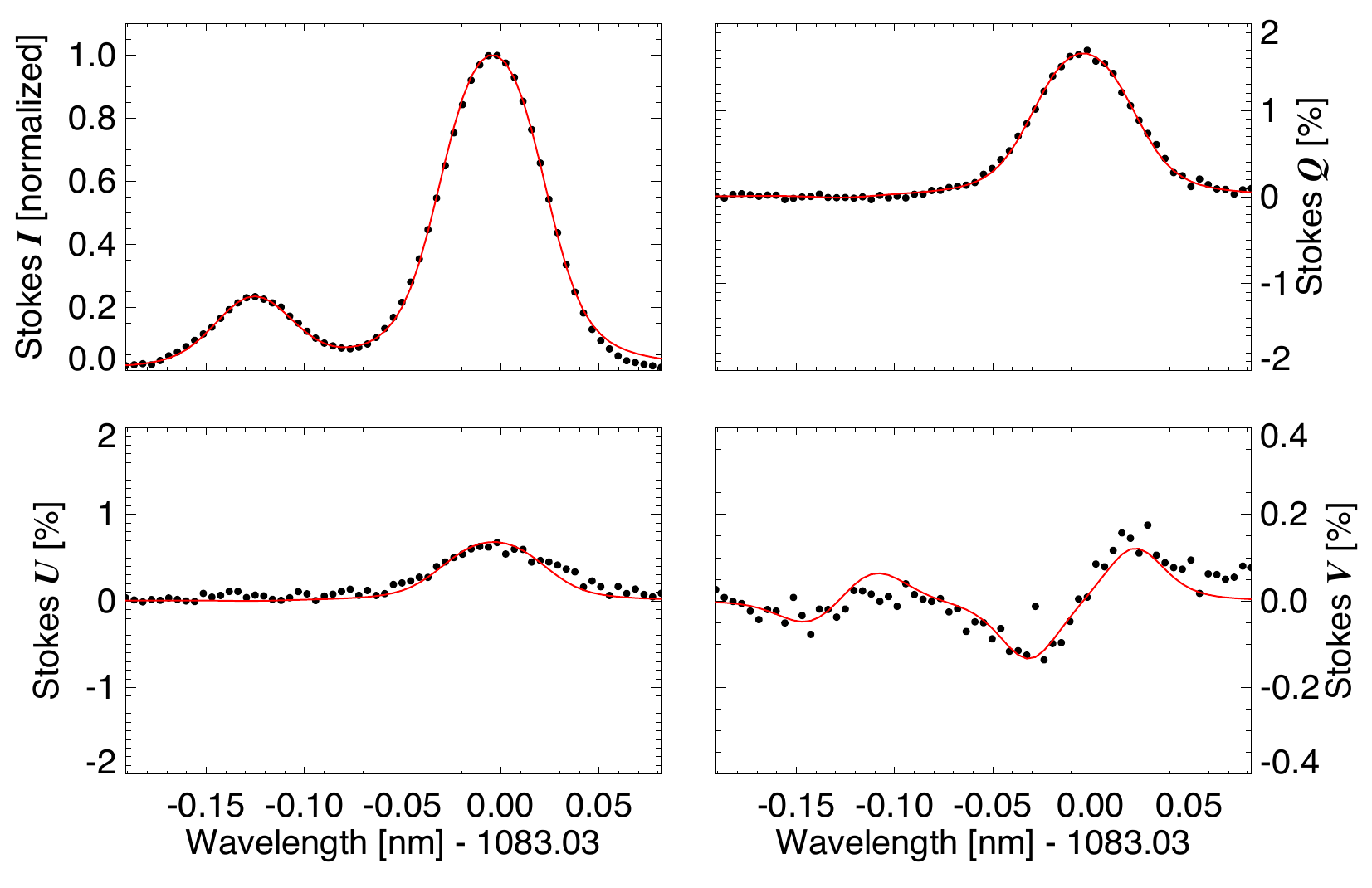}
\caption{\label{fig2} Example of the He~{\small I}~1083.0~nm Stokes profiles observed in the prominence. Dots represent the observations and the solid lines the theoretical profiles obtained by HAZEL. Stokes I is normalized to unity. Stokes Q, U, and V are normalized to their Stokes I maximum peak value. The inferred magnetic field vector has B~$\approx$~11~G, $\theta_\mathrm{B}\approx$ 70$^\mathrm{o}$, and $\chi_\mathrm{B}\approx$ 50$^\mathrm{o}$. }
\end{figure}

The observations were taken on 20 May 2011 with the Tenerife Infrared Polarimeter (TIP-II; \cite{2007ASPC..368..611C}) installed at the German Vacuum Tower Telescope (VTT) at the Observatorio del Teide (Tenerife, Spain). The VTT spectrograph scanned a region of about 40$^{\prime\prime}$ containing a quiescent, hedgerow prominence located in the  solar south east limb. The four Stokes parameters were measured with TIP-II around the 1083.0~nm spectral region containing the He~{\small I} multiplet as well as a Si~{\small I}~1082.70~nm photospheric line and an atmospheric water vapor line at 1083.21~nm. The spectral sampling was 1.1~pm. The length of the spectrograph's slit was 80$^{\prime\prime}$ and the scanning step 0$.\!\!^{\prime\prime}$5, which provided us with a 80$^{\prime\prime}\times$40$^{\prime\prime}$ map. Thanks to the adaptive optics system and the good seeing conditions during the observing run we measured the four Stokes parameters produced by the prominence plasma with a resolution of about 1$^{\prime\prime}$. The exposure time per polarization state was 15 seconds. The TIP-II data reduction process included dark current, flat-field, and fringes correction as well as the polarimetric calibration. 

\begin{figure}[t]
\center
\includegraphics[scale=0.6]{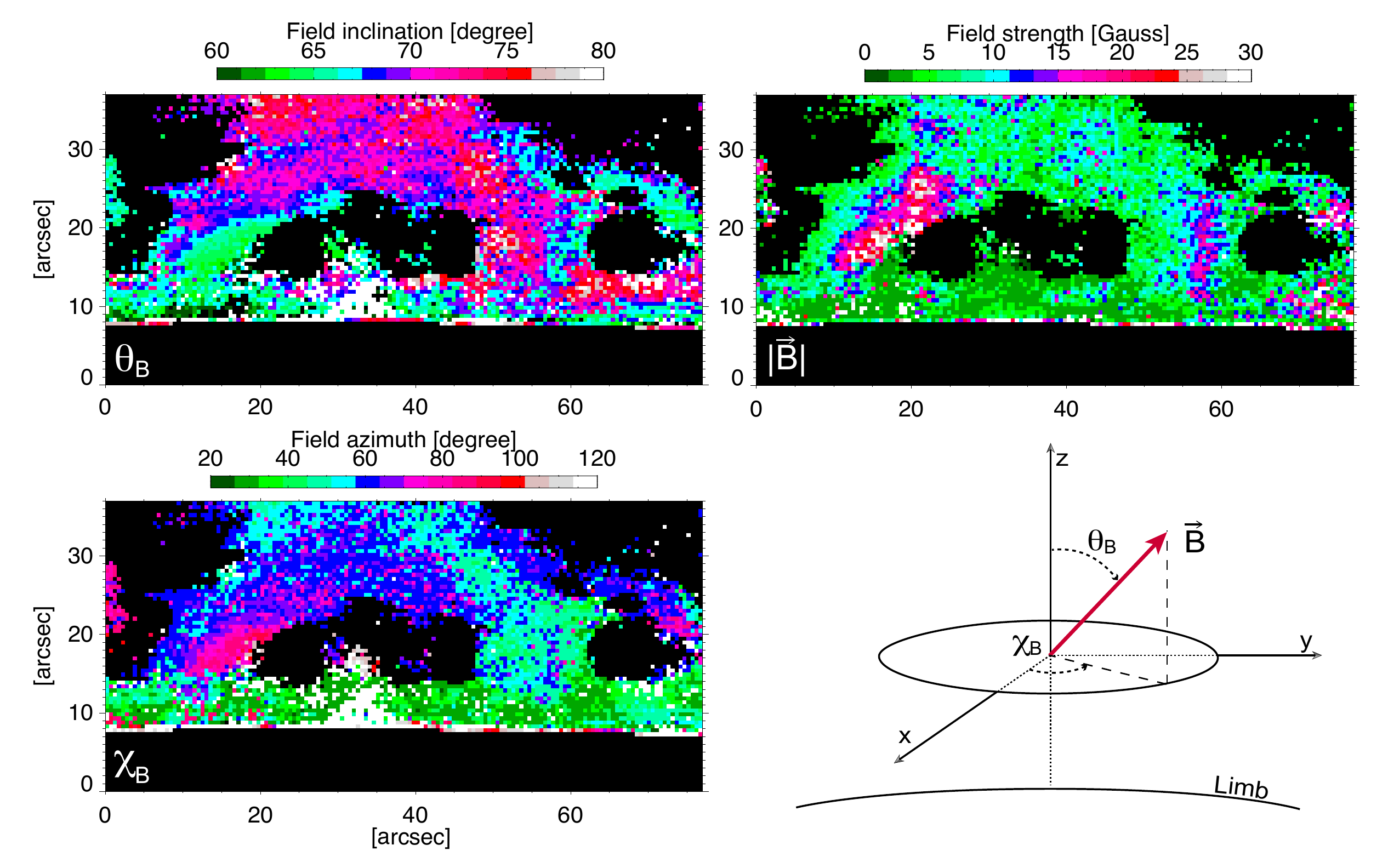}
\caption{\label{fig3} Field strength, inclination, and azimuth maps inferred from the interpretation of the observed Stokes profiles with the HAZEL code. As in Figure\ref{fig1}, the bottom region in all panels represent the solar limb. The rest of dark areas correspond to pixels whose Stokes Q or U signals did not exceed 5 times their corresponding noise levels. The bottom left panels represents the geometry of the reference system. The inclination $\theta$ is measured with respect to the solar local vertical Z. The azimuth is measured from the line-of-sight direction X.}
\end{figure}

Figure~\ref{fig1} shows the peak intensity observed in the He~{\small I}~1083.0~nm line. The prominence can be seen as a filamentary, loop-like structure. Figure~\ref{fig2} displays a prototypical He~{\small I}~1083.0~nm emission profile pertaining to the prominence. The Stokes I profile shows the fine structure of the triplet, with a weak blue component at 1082.9~nm separated about 0.12~nm from the other two, blended  components at about 1083.03~nm. The polarization signals illustrate nicely the joint action of the Hanle and Zeeman effect in the He~{\small I}~1083.0~nm triplet: Stokes Q and U shapes are dominated by atomic level polarization and the Hanle effect in a optically thin media\footnote{The typical signature of the presence of atomic polarization in 90$^\mathrm{o}$ scattering geometry is a nonzero Stokes Q profile. According
to the Hanle effect, the presence of a magnetic field inclined with respect to the local vertical direction produces the Stokes U signal. Note that, in the case of an optically thin media, only the red component of the triplet shows significant linear polarization signal\cite{2002Natur.415..403T}.} while Stokes V shows the typical circular polarization profiles dominated by the Zeeman effect.

To infer the magnetic field vector in the prominence we have to interpret the shapes of the Stokes profiles taking into account the joint action of atomic level polarization and the Hanle and Zeeman effects, as well as a suitable model to do the radiative transfer calculations. To this end, we applied the HAZEL code\cite{2008ApJ...683..542A} assuming a constant-property slab, which requires the determination of only seven parameters: the optical depth of the red component of the triplet, the line damping, the thermal velocity, the bulk velocity of the plasma, and the strength B ,inclination $\theta_\mathrm{B}$, and azimuth $\chi_\mathrm{B}$ of the magnetic field vector. The inference strategy in HAZEL consists on the comparison of the observed Stokes profiles with synthetic guesses of the signals. This is done by combining the Levenberg-Marquardt minimization method with a global optimization method, to avoid getting trapped in local minima of the merit function hypersurface. The only parameter we keep constant in the inversion is the height above the limb of each prominence point, which fixes the degree of anisotropy of the incident radiation field.

Figure~\ref{fig3} shows the distribution of the field strength, inclination, and azimuth across the prominence obtained from the analysis of the Stokes profiles with the HAZEL code. Only pixels whose linear polarization signals exceeded five times the noise level have been inverted. The analysis of the prominence data gave a typical field strength of about 5~G. Interestingly, we do not detect abrupt changes in the field strength along the prominence, contrary to the results of Casini et al.\ (2003). We only appreciate a stronger concentration of field in the left part of the prominence with field strength values up to 30~Gauss. Note that for fields stronger than 10~G it is only possible to determine the orientation of the field vector when using only the Hanle effect as a diagnostic tool. The reason is that for B~$\gtrsim$10~G the He~{\small I}~1083.0~nm line enters in the so-called Hanle saturation regime. Thus, in order to infer the field strength in those pixels we need to measure Stokes V signals as well. In our case, the signal to noise was good enough to detect not only Stokes Q and U signals above the noise, but also Stokes V. The profiles displayed in Figure~\ref{fig2} correspond to one of the pixels showing measurable linear and circular polarization signals. For fields below 10~G the Hanle effect also provides information on the field strength. 

Regarding the inclination of the field, we find that the magnetic field vector is almost horizontal with respect to the solar surface and that the inclination varies from about 65$^\mathrm{o}$ at the left part of the prominence to 75$^\mathrm{o}$ at the right side. Finally, the azimuth also varies along the prominence, being about 80$^\mathrm{o}$ at the left side of the prominence and 40$^\mathrm{o}$ at the right side. 

The magnetic configuration we show in Figure~\ref{fig3} is the most probable one, after having made a detailed study of the Van Vleck ambiguity of the Hanle effect. For resolving the possible ambiguities in the magnetic field orientation we made use of context information coming from the extreme ultraviolet light telescope onboard NASA's Solar Dynamics Observatory STEREO-B EUVI observations. For details we refer the reader to \cite{inprep}.

Note that the magnetic field map shown in Figure~\ref{fig3} includes the observed prominence as well as chromospheric spicules. The spicules can be seen right above the limb at about 8$^{\prime\prime}$--14$^{\prime\prime}$. Note also that the spicules region includes the contribution of both structures, i.e., prominence and spicules, and thus the interpretation of the inversion results regarding spicules is not straightforward in this data set. 

\section{Conclusions}

In this contribution we have shown that spectropolarimetric observations in the He~{\small I}~1083.0~nm triplet can be used to determine maps of the strength and orientation of the magnetic field in solar prominences. Here we have applied an inversion code that includes all necessary physics for interpreting the Stokes I, Q, U, and V profiles of the He~{\small I}~1083.0~nm triplet (HAZEL), what makes the analysis of the observations relatively straightforward. 

We have shown a case study in which we observed a quiescent, hedgerow prominence with the TIP-II instrument mounted on the VTT and then applied the HAZEL inversion code to the spectropolarimetric data. Our results show that the magnetic field is tilted about 10$^\mathrm{o}$--50$^\mathrm{o}$ with respect the solar limb direction and that it is inclined by 65$^\mathrm{o}$--75$^\mathrm{o}$ with respect to the local vertical, with field strength values between 5~G and 30~G. These results are in agreement with the known global magnetic properties of solar prominences. Interestingly, we did not find local enhancements in the field strength as those described in \cite{casini}. 

%
%
\small  
%
\section*{Acknowledgments}   
%
Financial support by the Spanish Ministry of Economy and Competitiveness (MINECO) through the project AYA2010-18029 (Solar Magnetism and Astrophysical Spectropolarimetry) is gratefully acknowledged. AAR also acknowledges financial support through the Ram\'on y Cajal fellowship.
%

%
\end{document}